\documentstyle[twoside,fleqn,espcrc2,epsf]{article}

\title{
Continuum Limit of Scalar Masses and Mixing Energies}

\author{W.\ Lee \thanks{present address:
T-8, LANL, Los Alamos, NM 87545}
and D.\ Weingarten\\
IBM Research, P.O.~Box 218,
Yorktown Heights, NY 10598, USA\\}

\begin{document}

\begin{abstract}

We evaluate the continuum limit of the valence approximation to the mass
of scalar quarkonium and to the scalar quarkonium-glueball mixing energy
for a range of different quark masses. Our results answer several
questions raised by the proposed identification of $f_0(1710)$ as
composed primarily of the lightest scalar glueball.
 
\end{abstract}

\maketitle

Evidence that $f_0(1710)$ is composed mainly of the lightest scalar
glueball is now given by a calculation~\cite{Sexton95}, on a $16^3
\times 24$ lattice at $\beta$ of 5.70, yielding 108(29) MeV as
the width for the lightest scalar glueball to decay to all possible
pseudoscalar pairs, and by three independent
calculations~\cite{Vaccarino,otherglue} with a combined
prediction~\cite{latestglue} of 1632(49) MeV as the infinite volume,
continuum limit of the lightest scalar glueball mass. All four
calculations were done in the valence (quenched) approximation.  The
decay width result combined with any reasonable guesses for the effect
of finite lattice spacing, finite lattice volume, and the remaining
width to multibody states yields a total width small enough for the
lightest scalar glueball to be seen easily in experiment. The mass
prediction combined with the expectation~\cite{Weingarten97} that the
valence approximation will underestimate the scalar glueball mass then
points to $f_0(1710)$ as composed primarily of the lightest scalar
glueball~\cite{Sexton95}.

Among established resonances the only plausible alternative to
$f_0(1710)$ is $f_0(1500)$.  Refs.~\cite{Sexton95,Lee97} propose that
$f_0(1500)$ consists mainly of $s\overline{s}$ scalar quarkonium. A
problem with this suggestion, however, is that $f_0(1500)$ apparently
does not decay mainly into states containing an $s$ and an
$\overline{s}$ quark~\cite{Amsler1}.  Ref.~\cite{Amsler2} thus
interprets $f_0(1500)$ as the lightest scalar glueball and $f_0(1710)$
as $s\overline{s}$ scalar quarkonium.  A further problem for any of
these identifications of $f_0(1500)$ and $f_0(1710)$ is that the
Hamiltonian of full QCD couples quarkonium and glueballs so that
physical states should be linear combinations of both.  In the extreme,
mixing could lead to $f_0(1710)$ and $f_0(1500)$ each half glueball and
half quarkonium.

For a fixed lattice period $L$ of about 1.6 fm and several different
values of quark mass, we have now found the valence approximation to the
continuum limit of the mass of the lightest $q\overline{q}$ states and
of the mixing energy between these states and the lightest scalar
glueball. Continuum predictions are extrapolated from calculations at
four different lattice spacings. For the two largest lattice spacings we
have results also for lattice period 2.3 fm.  Earlier stages of this
work are reported in Refs.~\cite{Lee97,Weingarten97,Lee98a}.

Our results provide answers to the questions raised by the
identification of $f_0(1710)$ as composed largely of the lightest scalar
glueball. With $L$ of 1.6 fm, the valence approximation to the continuum
limit of the mass of the lightest $s\overline{s}$ scalar we find is
significantly below the valence approximation to the infinite volume,
continuum limit of the scalar glueball mass. Our calculations with $L$
of 2.3 fm show that as the infinite volume limit is taken, the mass of
the lightest $s\overline{s}$ scalar will fall still further. Thus it
appears to us that the identification of $f_0(1500)$ as mainly glueball
with $f_0(1710)$ as mainly $s\overline{s}$ is quite improbable.  Our values
for mixing energy combined with the simplification of considering mixing
only among the lightest discrete isosinglet scalar states, then support
a set of physical mixed states with $f_0(1710)$ composed of 73.8(9.5)\%
glueball and $f_0(1500)$ consisting of 98.4(1.4)\% quarkonium, mainly
$s\overline{s}$. The glueball amplitude which leaks from $f_0(1710)$
goes almost entirely to the state $f_0(1390)$, which remains mainly
$n\overline{n}$, normal-antinormal, the abbreviation we adopt for
$(u\overline{u} + d\overline{d})/\sqrt{2}$. In addition, $f_0(1500)$
acquires an $n\overline{n}$ amplitude with sign opposite to its
$s\overline{s}$ component. Interference between the $n\overline{n}$ and
$s\overline{s}$ components of $f_0(1500)$ suppresses the state's decay
to $K\overline{K}$ final states by a factor consistent, within
uncertainties, with the experimentally observed suppression.

An alternative calculation of mixing between valence approximation
quarkonium and glueball states is given in Ref.~\cite{Boglione}.  We
show in Ref.~\cite{Lee98c}, however, that the calculation of
Ref.~\cite{Boglione} is not correct.

Our calculations, using Wilson fermions and the plaquette action, were
done with ensembles of 2749 configurations on a lattice $12^2 \times 10
\times 24$ with $\beta$ of 5.70, 1972 configurations on $16^3 \times 24$ with
$\beta$ of 5.70, 2328 configurations on $ 16^2 \times 14 \times 20$ with
$\beta$ of 5.93, 1733 configurations on $24^4$ at $\beta$ of 5.93, 1000
configurations on $24^2 \times 20
\times 32$ with $\beta$ of 6.17, and 1003 configurations on $32^2 \times
28 \times 40$ with $\beta$ of 6.40.  The smaller lattices with $\beta$
of 5.70 and 5.93 and the lattices with $\beta$ of 6.17 and 6.40 have
periods in the two (or three) equal space directions of 1.68(5) fm,
1.54(4) fm, 1.74(5) fm, 1.66(5) fm, respectively, permitting
extrapolations to zero lattice spacing with nearly constant physical
volume.  Conversions from lattice to physical units in this paper are
made~\cite{Vaccarino,latestglue} using the exact solution to the
two-loop zero-flavor Callan-Symanzik equation for
$\Lambda^{(0)}_{\overline{MS}} a$ with $\Lambda^{(0)}_{\overline{MS}}$
of 234.9(6.2) MeV~\cite{Butler}.

\begin{figure}
\epsfxsize=63mm
\epsfbox{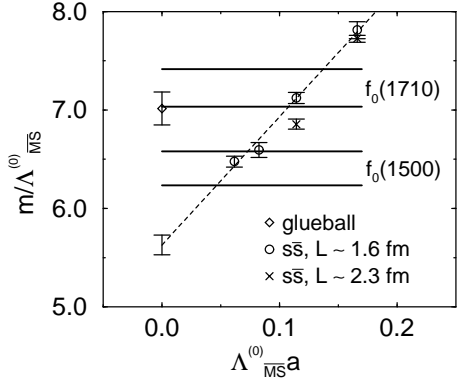}
\vskip -12mm
\caption{
Continuum limit of the scalar
$s\overline{s}$ mass, the scalar glueball mass, and one
sigma upper and lower bounds on observed masses.}
\vskip -10mm
\label{fig:masscont}
\end{figure}

From each ensemble of configurations, we evaluated, following
Ref.~\cite{Vaccarino,Lee97,Lee98a}, the lightest pseudoscalar quarkonium
mass, scalar quarkonium mass, scalar glueball mass and
quarkonium-glueball mixing energy $E$. These calculations were done for
a range of quark masses starting a bit below the strange quark mass and
running to a bit above twice the strange quark mass.

For $L$ near 1.6 fm, Figure~\ref{fig:masscont} shows the $s\overline{s}$
scalar mass in units of $\Lambda^{(0)}_{\overline{MS}}$ as a function of
lattice spacing in units of $1/\Lambda^{(0)}_{\overline{MS}}$.  A linear
extrapolation of the mass to zero lattice spacing gives 1322(42) MeV,
far below our valence approximation infinite volume continuum glueball
mass of 1632(49) MeV.  Figure~\ref{fig:masscont} shows also values of
the $s\overline{s}$ scalar mass at $\beta$ of 5.70 and 5.93 with $L$ of
2.24(7) and 2.31(6) fm, respectively.  The $s\overline{s}$ mass with $L$
near 2.3 fm lies below the 1.6 fm result for both values of lattice
spacing. Thus the infinite volume continuum $s\overline{s}$ mass should
lie below 1322(42) MeV.  For comparison with our data,
Figure~\ref{fig:masscont} shows the valence approximation value for the
infinite volume continuum limit of the scalar glueball mass and the
observed value of the mass of $f_0(1500)$ and of the mass of
$f_0(1710)$.

\begin{figure}
\epsfxsize=63mm
\epsfbox{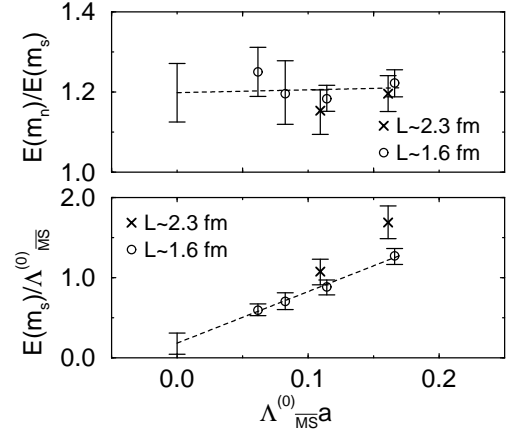}
\vskip -10mm
\caption{Continuum limit of the
mixing energy $E(m_s)$ and of the ratio $E(m_n)/E(m_s)$.}
\vskip -10mm
\label{fig:mixcont}
\end{figure}

Figure~\ref{fig:mixcont} shows a linear extrapolation to zero lattice
spacing of quarkonium-glueball mixing energy at the strange quark mass
$E(m_s)$. The extrapolation uses the points with $L$ near 1.6 fm. The
points with larger lattice period suggest that $E(m_s)$ rises a bit
with lattice volume, but the trend is not statistically significant.  As
a function of quark mass with lattice spacing fixed, we found the mixing
energy to be extremely close to linear. We were thereby able to
extrapolate our data reliably down to the normal quark mass $m_n$.
Figure~\ref{fig:mixcont} shows also a linear extrapolation to zero
lattice spacing of the ratio $E(m_n)/E(m_s)$. The fit is to the set of
points with $L$ near 1.6 fm but is also consistent with the
points for larger $L$. Thus the continuum limit we obtain for
$E(m_n)/E(m_s)$ is also the infinite volume limit. The limiting value of
$E(m_s)$ is 43(31) MeV and of $E(m_n)/E(m_s)$ is 1.198(72).

The infinite volume continuum value for $E(m_n)/E(m_s)$ we now take as
an input to a simplified treatment~\cite{Weingarten97} of the mixing
among valence approximation glueball and quarkonium states which arises
in full QCD from quark-antiquark annihilation.  The Hamiltonian
coupling together the scalar glueball, the scalar $s\overline{s}$ and
the scalar $n\overline{n}$ isosinglet is
\begin{displaymath}
\left| 
\begin{array}{ccc}
m_g &  E(m_s) & \sqrt{2} r E(m_s) \\
E(m_s) & m_{s\overline{s}} & 0 \\
\sqrt{2} r E(m_s) & 0 & m_{n\overline{n}}. 
\end{array} 
\right| 
\end{displaymath} 
where $r$ is $E(m_n)/E(m_s)$, and $m_g$, $m_{s\overline{s}}$
and $m_{n\overline{n}}$ are, respectively, the glueball mass, the
$s\overline{s}$ quarkonium mass and the $n\overline{n}$ quarkonium mass
before mixing.

The three unmixed mass parameters and $E(m_s)$, for which our measured
value has a large fractional error bar, we determine from four observed
masses. To leading order in the valence approximation, with valence
quark-antiquark annihilation turned off, corresponding isotriplet and
isosinglet states composed of $u$ and $d$ quarks will be degenerate. For
$m_{n\overline{n}}$ we thus take the observed isovector value of
1470(25) MeV~\cite{Amsler1}. The three remaining unknowns we tune to give
the mixing Hamiltonian eigenvalues of 1697(4) MeV, 1505(9) MeV and
1404(24) MeV, respectively the Particle Data Group's masses for
$f_0(1710)$ and $f_0(1500)$, and the weighted average of
Refs.~\cite{Amsler1,MarkIII} masses for $f_0(1390)$.

We find $m_g$ becomes 1622(29) MeV, $m_{s\overline{s}}$ becomes
1514(11) MeV, and $E(m_s)$ becomes 64(13) MeV, with error bars including
the uncertainties in the four input physical masses. The unmixed $m_g$
is consistent with the world average valence approximation glueball mass
1632(49) MeV, $E(m_s)$ is consistent with our
measured value of 43(31) MeV, and $m_{s\overline{s}}$ is about 13\% above the valence
approximation value 1322(42) MeV for lattice period 1.6 fm.  This 13\%
gap is comparable to the largest disagreement, about 10\%, found between
the valence approximation and experimental values for the masses of
light hadrons.  In addition, the physical mixed $f_0(1710)$ has a
glueball content of 73.8(9.5)\%, the mixed $f_0(1500)$ has a glueball
content of 1.6(1.4)\% and the mixed $f_0(1390)$ has a glueball content
of 24.5(10.7)\%.  These predictions are supported by a recent reanalysis
of Mark III data~\cite{MarkIII} for $J/\Psi$ radiative decays.  Finally,
the state vector for $f_0(1500)$ we find has a relative negative sign
between the $s\overline{s}$ and $n\overline{n}$ components leading, by
interference, to a suppression of the partial width for this state to
decay to $K\overline{K}$ by a factor of 0.39(16) in comparison to the
$K\overline{K}$ rate for an unmixed $s\overline{s}$ state. This
suppression is consistent with the experimentally
observed suppression.

\end{document}